\documentclass[published]{JHEP3} % 10pt is ignored!

\JHEP{00(2007)000}

\JHEPspecialurl{http://jhep.sissa.it/JOURNAL/JHEP3.tar.gz}

\usepackage{epsfig,multicol,bbm}

\newcommand\fverb{\setbox\fverbbox=\hbox\bgroup\verb}
\newcommand\fverbdo{\egroup\medskip\noindent%
			\fbox{\unhbox\fverbbox}\ }
\newcommand\fverbit{\egroup\item[\fbox{\unhbox\fverbbox}]}
\newbox\fverbbox

\def\bequ{\begin{equation}}
\def\eequ{\end{equation}}
\def\barr{\begin{array}}
\def\earr{\end{array}}

\title{A Double Myers-Perry Black Hole in Five Dimensions}

\author{Carlos A. R. Herdeiro, Carmen
Rebelo,  Miguel Zilh\~ao, Miguel S. Costa
\\
Departamento de F\'\i sica e Centro de F\'\i sica do Porto \\
Faculdade de Ci\^encias da
Universidade do Porto \\
Rua do Campo Alegre, 687,  4169-007 Porto, Portugal\\
	E-mails:
	\email{crherdei,mrebelo,miguelc@fc.up.pt;c0302009@alunos.fc.up.pt (MZ)}}

\received{May 24, 2008} 		%%
\accepted{June 21, 2008}		%% These are for published papers.

\preprint{\hepth{9912999}}	% OR: \preprint{Aaaa/Mm/Yy\\Aaa-aa/Nnnnnn}
\abstract{Using the inverse scattering method we construct a six-parameter
family of exact, stationary,
asymptotically flat solutions of the 4+1 dimensional vacuum Einstein
equations, with $U(1)^2$ rotational symmetry. It describes the
superposition of two Myers-Perry black holes, each with a
\textit{single} angular momentum parameter, both in the same
plane. The black holes live in a background geometry which is
the Euclidean C-metric with an extra flat time direction. 
This background possesses conical singularities in
two adjacent compact regions, each corresponding to a set of fixed
points of one of the $U(1)$ actions in the Cartan sub-algebra of
$SO(4)$. We discuss several aspects of the black holes geometry, including the conical singularities arising from force imbalance, and the torsion singularity arising from torque imbalance. 
The double Myers-Perry solution presented herein
is considerably simpler than the four dimensional double Kerr solution
and might be of interest in studying spin-spin interactions in five 
dimensional general relativity.}


\begin{document} 

%\maketitle  IS IGNORED %%%%%%%%%%%

\section{Introduction}
The paradigmatic example of a static, regular (on and outside an event
horizon) multi-black hole spacetime is the family of
Majumdar-Papapetrou solutions of Einstein-Maxwell theory in four
dimensions \cite{Majumdar:1947eu,Papapetrou}. All of the individual objects
in these configurations are extremal Reissner-Nordstr\"om black holes
\cite{Hartle:1972ya}, which are held in equilibrium due to a
balance between gravitational attraction and electrostatic repulsion
for any pair of black holes. Such force balance is
mathematically realised by an \textit{exact} linearisation of the full
Einstein-Maxwell equations. This linearisation is most easily obtained
by taking the  Einstein-Maxwell theory as the bosonic sector of
$\mathcal{N}=2$, $D=4$ Supergravity and searching for static,
supersymmetric backgrounds with a timelike Killing vector field; the
Majumdar-Papapetrou family is the most general such solution
\cite{Gibbons:1982fy}.

It turns out that the Majumdar-Papapetrou family is not the
most general \textit{stationary}, supersymmetric background within
$\mathcal{N}=2$, $D=4$ Supergravity, even demanding asymptotic
flatness \cite{Tod:1983pm}; the most general such solution is the
Israel-Wilson-Perjes (IWP) family \cite{Israel:1972vx,Perjes:1971gv}. 
For a specific choice, it
represents a set of Kerr-Newman ``particles'' (naked singularities),
each of which is obtained by giving spin to an extremal
Reissner-Nordstr\"om black hole. The force balance is now more
involved: in addition to the monopole-monopole gravitational
attraction and electrostatic repulsion, we have dipole-dipole 
forces. The gravitational one is a spin-spin force, first discussed by
Wald \cite{Wald:1972sz} using Papapetrou's equation for a spinning
particle \cite{Papapetrou:1951pa}. Wald showed that, in an appropriate
limit, this force has exactly the same form as the usual dipole-dipole
force in magnetostatics, but with \textit{opposite
sign}\footnote{Actually, using a gravito-electromagnetic analogy 
based on tidal tensors \cite{FilipeCosta:2006fz}, the Papapetrou equation 
for a spinning particle can be simply derived from the force
acting on a magnetic dipole in magnetostatics.}.
This fact clarifies why there is a force balance in the IWP spacetimes, 
independently of the orientation of the spin of the
individual black holes. This is furthermore confirmed by a probe
computation for a charged spinning particle in an IWP spacetime
\cite{Kastor:1998cy} and by a post-post Newtonian analysis of 
the metric generated by two massive
charged spinning sources in the Einstein-Maxwell theory \cite{Bonnor:2001mh}. 
Note that, since the magnetic dipole of a
charged spinning black hole is not an independent quantity, the
gyromagnetic ratio plays a crucial role in the cancellation of
dipole forces. 

But the balance of forces does not guarantee equilibrium in the presence of dipoles. 
We also have to discuss the balance of \textit{torques}, which is more subtle. 
Like in magnetostatics, in general relativity
non-aligned gravitational dipoles (spinning bodies) also produce 
a torque on each other \cite{Schiff:1960gi}, which has been
recently tested by the Gravity Probe B experiment. This torque
obviously vanishes when the two spins are aligned, but \textit{not
the total torque}. Imagine that a Schwarzschild black hole is placed
in the vicinity of a Kerr black hole, with the spin of the latter
parallel to the direction of separation. One could impose a
constraint (in the form of a strut) preventing the two black holes
from approaching, i.e. from gaining linear acceleration. If no
constraint is imposed in the form of a torque, we would expect the
Schwarzschild black hole to gain angular acceleration, due to the
dragging of inertial frames caused by the Kerr black hole. Thus, in
the gravitational case, there seems to be an additional torque,
besides the aforementioned one.

If this additional torque is present  we might expect some signature
in a multi black hole spacetime. Indeed, it was shown in
\cite{bonnorward} that the rotation one-form in a two (aligned)
particles IWP spacetime  will diverge somewhere along the axis - either in between
the particles or in the remaining of the axis - unless a certain
requirement, which we dub \textit{axis condition}, is obeyed
(c.f. section \ref{axissection}). Failure
to obey this condition has been interpreted as a ``torsion
singularity''  in \cite{Bonnor:2001mh,Bonnor:2001,Letelier:1998ft};
therein the condition arises as the requirement that the azimuthal
vector field  has a fixed point at the axis and is spatial
otherwise. The analysis in \cite{Bonnor:2001mh} also suggests that,
for charged rotating black holes, there is
an electromagnetic contribution to the effect that makes
a Schwarzschild black hole rotate in the vicinity of a Kerr black
hole. However, it so happens that for this effect the
purely gravitational and electromagnetic contributions do not completely
cancel, even in the supersymmetric case of IWP particles. 
It is worth noting that the post-post Newtonian analysis suggests that, for
uncharged sources, the \textit{regularity condition}, i.e. the requirement 
of absence of conical singularity representing struts or strings necessary for
force balance, is incompatible with the axis condition \cite{Bonnor:2001}. 

The regularity condition has been studied at the level of exact,
non-supersymmetric, static solutions in the multi-Schwarzschild
\cite{israelkhan,Costa:2000kf} and
the multi-Schwarzschild-Tangherlini \cite{Tan:2003jz} spacetimes. 
However, to study the
regularity and axis condition at the level of exact, non-supersymmetric 
solutions seems, in principle, a much more
difficult task, mainly because such solutions, which are stationary
rather than static, are usually rather
involved. The paradigmatic example is the double-Kerr solution,
originally generated in \cite{Kramer:79} via a B\"acklund
transformation. The complexity of this solution has led to different claims
concerning the explicit form of these conditions (see, for instance
\cite{Dietz,Letelier:1998ft,Manko:2002}), although it is a consensual
conclusion that the solution for two black holes must have
singularities, in agreement with the spinning test particle analysis
of \cite{Wald:1972sz}.  It turns out that a
five dimensional version of the double-Kerr solution, the
\textit{double Myers-Perry} solution, is drastically simpler than its
four dimensional counterpart. The reason is simply understood: using
the Belinskii-Zakharov inverse scattering method
\cite{Belinskii:78,Belinskii:79}, the Kerr solution \cite{Kerr:1963ud}
is generated by a
2-soliton transformation, whereas the Myers-Perry solution
\cite{Myers:1986un} with a
single angular momentum can be generated by a \textit{single}
soliton transformation (see \cite{Emparan:2008eg} for a review of the inverse scattering
method and applications). Thus, whereas the double-Kerr 
solution is generated by a 4-soliton transformation
\cite{Letelier:1998ft}, the double-Myers-Perry solution is generated,
effectively, by a 2-soliton transformation\footnote{Actually we will
  use a 4-soliton transformation, but since two of the solitons will
  have trivial BZ vectors, it is effectively as complex as a 2-soliton
  transformation.}. To generate the latter
solution is the main purpose of the present paper. This will allow us
to write down in a very simple and clear fashion the regularity and axis
conditions for this spacetime.

The new solution presented herein is also of interest in a different
context. Over the last few years a great effort has been made to tackle
the black hole classification problem in higher dimensions
\cite{Emparan:2008eg}. It is well known that the ``phase space'' of
regular (i.e. free of curvature singularities on and outside an event horizon) and asymptotically flat black
objects is
rather richer than in four dimensions, containing exotic objects
such as black rings \cite{Emparan:2001wn,Mishima:2005id,Figueras:2005zp,Pomeransky:2006bd,Elvang:2007hs,Iguchi:2007is,Evslin:2007fv} and black 
saturns \cite{Elvang:2007rd}; equivalently there are no
(simple) black hole uniqueness theorems analogous to the four dimensional
case for vacuum, stationary configurations. 
The new stationary solution presented herein describes the
superposition of two Myers-Perry black holes in five dimensions, each with a
single angular momentum parameter, both in the same
plane. The black holes live in a background geometry, which is
the Euclidean C-metric with an extra flat time direction. The downside of the new solution is that it is built upon a non-trivial background geometry with conical singularities, which are still present, generically, when the black holes are included. It remains to be seen if, by including the second angular momentum parameter or other fields, like the electromagnetic field, such singularities can be removed.

This paper is organised as follows. In section two we analyse the
background geometry upon which the double Myers-Perry solution will be built. The use of this background is actually a necessity for using the
inverse scattering method. In section 3 we discuss the static
solution, first constructed in \cite{Tan:2003jz}, that will be used as
the seed metric for the new solution presented herein. In section 4 the
double Myers-Perry solution is generated using the inverse scattering
method and its rod structure is analysed; other basic properties of
the solution as well as the computation of the relevant physical
quantities are presented in section 5. We close with a discussion.

\section{Background geometry: The Euclidean C-metric}
\label{background}
In principle one could have a double Myers-Perry solution in five dimensions 
that would reduce to flat space when the two black holes are removed. Such 
solution, however, could not have a $U(1)^2$ spatial isometry which,
together with time translations, yields the three commuting Killing
vector fields necessary to apply the inverse scattering method 
that we shall use to generate the new solution. Indeed, placing two separated point-like sources in five dimensional Minkowski 
spacetime reduces 
the spatial isometry to $SO(3)$. Introducing rotation breaks this symmetry group 
further; at most we end up with $SO(2)$. Thus, such solution could not be 
generated by the Weyl or inverse scattering techniques and such problem seems 
very difficult to approach \cite{Myers:1986rx}. 

To generate a solution with 
two black holes (with topologically $S^3$ horizons) with the Weyl and inverse 
scattering techniques, we need a $U(1)^2$ spatial isometry and hence a background 
with at least \textit{two} fixed points of the two $U(1)$ actions. Flat space has 
only one such point, as it is clear from its rod structure. A background with two 
such points would be the four dimensional Euclidean Schwarzschild with an added 
time direction. However, as it is clear from its rod structure, this background 
is not asymptotically flat. The black holes one can superimpose on this background 
live on Kaluza-Klein bubbles, and they have been constructed in 
\cite{Elvang:2002br,Tomizawa:2007mz,Iguchi:2007xs}. 

\begin{figure}[h!]
\begin{picture}(0,0)(0,0)
\put(60,40){$y=-1$}
\put(153,40){$y=-\frac{1}{2mA}$}
\put(120,8){$x=+1$}
\put(-2,62){$t$}
\put(-5,33){$\phi$}
\put(-5,2){$\psi$}
\put(111,-5){$a_1$}
\put(151,-5){$a_3$}
\put(190,-5){$a_5$}
\put(228,8){$x=-1$}
\end{picture}
\centering\epsfig{file=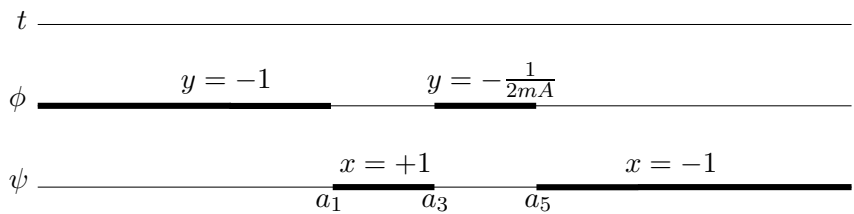,width=11cm}
\caption{Rod structure for the background spacetime. Next to each rod
  we write its locus in  $xy$ coordinates. The rods correspond to the
  edges of the rectangle in figure  2 (right). In terms of the parameters $m$ and $A$, the $a_i$'s can be taken as $a_1=-\frac{1}{2A^2}$, $a_3=-\frac{m}{A}$, $a_5=\frac{m}{A}$.}
\label{instantao}
\end{figure}

In order to have at least two fixed 
points and asymptotic flatness we need a background with three fixed points, 
which is exactly what happens for the Euclidean C-metric with an extra flat time 
direction; thus this geometry is our  background in the absence of the 
two black holes. Its rod structure is  represented in figure \ref{instantao}.

\begin{figure}[h!]
\begin{picture}(0,0)(0,0)
\put(320,49){$a_5$}
\put(418,49){$a_3$}
\put(310,79){$_{\partial/\partial \tilde{\psi}}$}
\put(420,79){$_{\partial/\partial \tilde{\psi}}$}
\put(365,39){$_{\partial/\partial \tilde{\phi}}$}
\put(365,118){$_{\partial/\partial \tilde{\phi}}$}
\put(418,112){$a_1$}
\end{picture}
\centering\epsfig{file=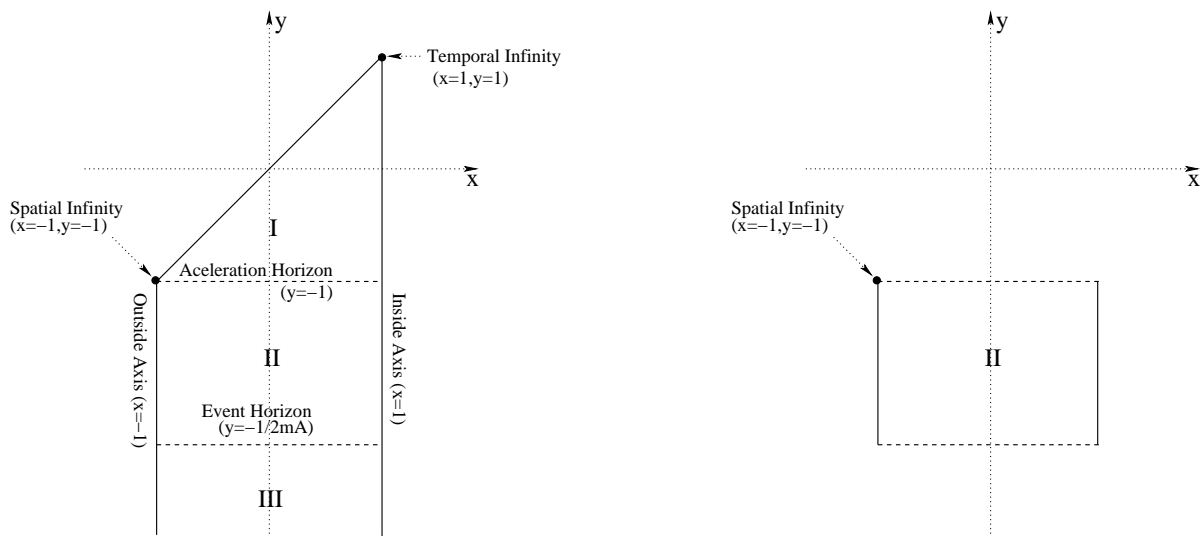,width=16cm}
\caption{$xy$ coordinate space for the Lorentzian (left) 
and Euclidean (right) C-metric. For the latter each edge is a 
fixed point set of a given periodic vector field, displayed next to it.}
\label{coordenadasxy}
\end{figure}

The \textit{Lorentzian} C-metric can be written 
in \textit{xy coordinates} as \cite{Hong:2003gx}
\[
ds^2=\frac{1}{A^2(x-y)^2}
\left[G(y)dt^2-\frac{dy^2}{G(y)}+\frac{dx^2}{G(x)}+G(x)d\tilde{\psi}^2\right] \ , \ \ \ \ 
G(\xi)\equiv(1-\xi^2)(1+2mA\xi) \ , 
\]
with $0<2mA<1$. The coordinate range for the $xy$ coordinates is displayed in 
figure \ref{coordenadasxy} (left). For the Lorentzian solution, these are 
$-1\le x\le 1$ for $-\infty<y\le-1$ and $-1\le y<x\le1$. The latter (region I) 
corresponds to a ``Milne'' region wherein the coordinate $t$ is spacelike and 
$y$ is timelike, a behaviour also found for these coordinates when $y<-1/2mA$ 
(region III).

The Euclidean C-metric we want to consider is obtained by the analytic 
continuation $t\rightarrow i\tilde{\phi}$. The coordinate range for the 
$xy$ coordinates is also displayed in figure \ref{coordenadasxy} (right), 
corresponding to the rectangular region wherein the $t$ coordinate was 
timelike in the Lorentzian regime (region II). The boundary of this region 
is a set of fixed points of either $\partial/\partial \tilde{\psi}$ at $x=\pm 1$, 
$\partial/\partial \tilde{\phi}$ at $y=-1,-\frac{1}{2mA}$, or both
(the three vertexes of the rectangle denoted $a_1$, $a_3$ and $a_5$ 
are \textit{double fixed points}). This can be clearly seen by changing from 
$(x,y)$ coordinates to canonical Weyl coordinates $(\rho,z)$; in particular we have 
\[
\rho^2=\frac{(y^2-1)(1-x^2)(1+2mAy)(1+2mAx)}{A^4(x-y)^4(1-2mA)^4} \ . 
\]
Note that the vertexes of the rectangular region in $xy$ coordinate space 
correspond to the breaks in the rod structure of the Euclidean 
C-metric - figure \ref{instantao}.

\begin{figure}[h!]
\begin{picture}(0,0)(0,0)
\put(150,60){Edge 1}
\put(150,50){Conical excess $\delta_{\psi}$}
\put(150,40){$\phi$-plane}
\put(35,30){Edge 2}
\put(35,20){Conical excess $\delta_{\phi}$}
\put(35,10){$\psi$-plane}
\end{picture}
\centering\epsfig{file=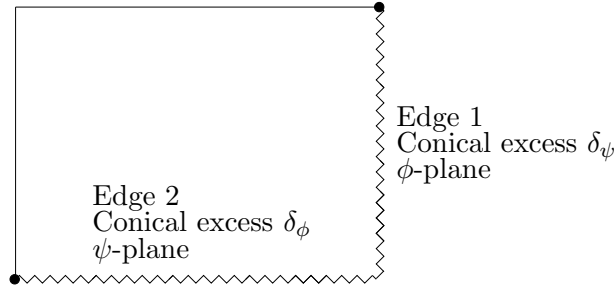,width=5cm}
\caption{Conical singularities in $xy$ coordinate space. After an appropriate choice of $\phi,\psi$ angles, with canonical period $2\pi$ the only conical singularities are found at the edges $x=+1$ (Edge 1), $y=-\frac{1}{2mA}$ (Edge 2). The two double fixed points where black holes will be placed are also emphasised.}
\label{conicalsingularities}
\end{figure}

\begin{figure}[h!]
\begin{picture}(0,0)(0,0)
\put(115,140){$||{\partial/\partial \phi}||$}
\put(340,140){$||{\partial/\partial \psi}||$}
\end{picture}
\centering\epsfig{file=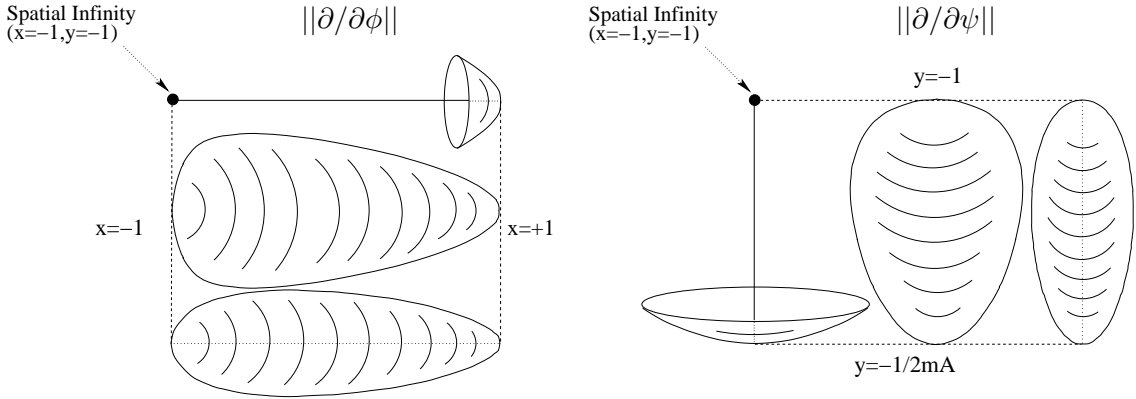,width=15cm}
\caption{$xy$ coordinate space; Left: Norm of $\frac{\partial}{\partial \psi}$ along $y=-\frac{1}{2mA},-\frac{1}{mA},-1$; note that $x=\pm 1$ are fixed point sets of this vector field, but with our choice of angular coordinate there are conical singularities only at $x=+1$;  Right: Norm of $\frac{\partial}{\partial \phi}$ along $x=-1,0,+1$; note that $y=-1,-\frac{1}{2mA}$ are fixed point sets of this vector field, but with our choice of angular coordinate there are conical singularities only at $y=-\frac{1}{2mA}$.}
\label{sizescircles}
\end{figure}

There are conical singularities in this background geometry. This is the price to pay to have three double fixed points. We can, however, make the geometry free of conical singularities at spatial infinity. Defining new angular coordinates $(\phi,\psi)\equiv (1-2mA)(\tilde{\phi},\tilde{\psi})$, and taking the canonical periodicities $\Delta \phi=2\pi=\Delta \psi$, the edges $x=-1$ and $y=-1$ become free of conical singularities. Thus the background geometry becomes asymptotically flat. In the remaining two edges there are conical \textit{excesses} - figure \ref{conicalsingularities} - given by
\[
\delta_{\psi}=2\pi\frac{4mA}{1-2mA}=2\pi\frac{a_{53}}{a_{31}} \ , \ \ \ \ \ 
x=+1 \ \Leftrightarrow \ a_1<z<a_3 \ , 
\]
\bequ
\delta_{\phi}=2\pi\frac{1-2mA}{4mA}=2\pi\frac{a_{31}}{a_{53}} \ , \ \ \ \ \ 
y=-\frac{1}{2mA} \ \Leftrightarrow \ a_3<z<a_5 \ , 
\label{cexy}
\eequ
where throughout this paper we use the notation
\[ a_{ij}\equiv a_i-a_j \ . \] 
In figure \ref{sizescircles} we represent the norm of $\partial/\partial \psi$ and $\partial/\partial \phi$ in $xy$ coordinate space. This gives an idea of the four dimensional Euclidean geometry. In particular, neglecting the conical singularities, its topology is $S^2\times S^2-\{P\}$, where the point $P$ corresponds to spatial infinity wherein these norms diverge. This topology is analogous to that of the instanton considered in \cite{Garfinkle:1990eq}.

In terms of canonical Weyl coordinates, the metric of the five dimensional background geometry has the form\footnote{Note that the dimensions of these coordinates are $[\rho]=[z]=L^2$.}
\bequ
ds^2=-dt^2+\frac{\mu_3}{\mu_1\mu_5}\rho^2 d\phi^2+\frac{\mu_1\mu_5}{\mu_3}\left(d\psi^2+\frac{(\rho^2+\mu_1\mu_3)^2(\rho^2+\mu_3\mu_5)^2\left[d\rho^2+dz^2\right]}{(\rho^2+\mu_1\mu_5)^2(\rho^2+\mu_1^2)(\rho^2+\mu_3^2)(\rho^2+\mu_5^2)}\right) \ , \eequ
where 
\[{\mu}_k\equiv \sqrt{\rho^2+(z-a_k)^2}-(z-a_k) \ . \]
This metric is invariant under the exchange
\bequ
a_1\leftrightarrow a_5 \ . \label{invarianceback}
\eequ
However notice that the conical excesses in the $\phi$ and $\psi$ plane are interchanged. Since a generalisation of this invariance will hold in the presence of
the two black holes, let us comment on it. The physical information that determines the geometry is given by the
sizes of the two finite rods in figure \ref{instantao}. Thus, one of
the three parameters that describe the geometry $(a_1,a_3,a_5)$ is
redundant. Such redundancy can be gauged away by introducing a new
coordinate $\tilde{z}=z-a_3$, in terms of which the metric reads
\[
ds^2=-dt^2+\frac{\mu}{\mu_{13}\mu_{53}}\rho^2 d\phi^2+\frac{\mu_{13}\mu_{53}}{\mu}\left(d\psi^2+\frac{(\rho^2+\mu\mu_{13})^2(\rho^2+\mu\mu_{53})^2\left[d\rho^2+d\tilde{z}^2\right]}{(\rho^2+\mu_{13}\mu_{53})^2(\rho^2+\mu_{13}^2)(\rho^2+\mu^2)(\rho^2+\mu_{53}^2)}\right) \ , \]
where 
\[{\mu}_{k3}\equiv \sqrt{\rho^2+(\tilde{z}-a_{k3})^2}-(\tilde{z}-{a}_{k3}) \ , \ \ \ \ \
\mu=\mu_{33} \ .\]
Fixing the physical information, i.e. the rod sizes $a_{53}$ and
$a_{31}$, it is simple to show that the metric is invariant under 
\[ (a_{31},a_{53};\tilde{z},\psi,\phi) \rightarrow
(a_{53},a_{31};-\tilde{z},\phi,\psi) \ . \]
This follows easily by noting that under this transformation 
\[ \mu_{53}\rightarrow \frac{\rho^2}{\mu_{13}} \ , \ \ \ \
\mu_{13}\rightarrow \frac{\rho^2}{\mu_{53}} \ , \ \ \ \ \mu\rightarrow
\frac{\rho^2}{\mu} \ . \]
This is nothing but the usual invariance of a system of particles on a
line under the inversion of the order together with a parity
transformation and it is what the transformation
(\ref{invarianceback}) effectively implements. Noting this invariance will be useful for checking our
solution and also for checking physical quantities that describe the whole
spacetime. Note that, in the particular case $a_{31}=a_{53}$,
the background geometry is invariant under $(\tilde{z},\psi,\phi)\rightarrow
(-\tilde{z},\phi,\psi)$; this is the five dimensional version of the
$\mathbb{Z}_2$ symmetry of, for instance, the equal mass
double-Schwarzschild solution.

\section{The static case: double Schwarzschild-Tangherlini}
\label{staticsolution}
\begin{figure}[h!]
\begin{picture}(0,0)(0,0)
\put(87,74){$_{(1,0,0)}$}
\put(201,74){$_{(1,0,0)}$}
\put(30,43){$_{(0,1,0)}$}
\put(163,43){$_{(0,1,0)}$}
\put(260,12){$_{(0,0,1)}$}
\put(126,12){$_{(0,0,1)}$}
\put(-2,62){$t$}
\put(-5,33){$\phi$}
\put(-5,2){$\psi$}
\put(73,-5){$a_1$}
\put(111,-5){$a_2$}
\put(151,-5){$a_3$}
\put(190,-5){$a_4$}
\put(228,-5){$a_5$}
\end{picture}
\centering\epsfig{file=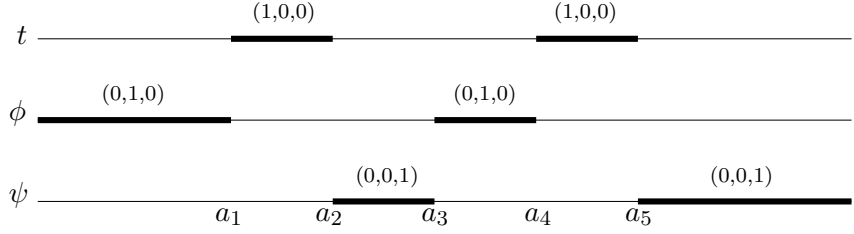,width=11cm}
\caption{Rod structure for the double Schwarzschild-Tangherlini. Next to each rod the corresponding eigenvector \cite{Harmark:2004rm} is displayed.}
\label{doubletan}
\end{figure}

The starting point for the new solution which will be presented in the 
next section is the double Schwarzschild-Tangherlini spacetime built 
in \cite{Tan:2003jz}, using the technique developed in
\cite{Emparan:2001wk}, whose rod structure is given in figure
\ref{doubletan}. We have placed the two timelike rods
representing black hole horizons at $z=a_1,a_5$ in figure
\ref{instantao}, so that the conical singularities represented in
figure \ref{conicalsingularities} are in between the two black
holes. In this way we expect that the interactions between the
two black holes might alter significantly these singularities. Note that, throughout this paper, we choose the ordering:
\bequ
a_1<a_2<a_3<a_4<a_5 \ . \label{ordering} \eequ

%\spa{0.05cm}

In the static case the metric is essentially read off from the rod structure:
\bequ
ds^2=-\frac{\mu_1 \mu_4}{\mu_2 \mu_5}dt^2+\frac{\mu_3}{\mu_1\mu_4}\rho^2 d\phi^2+\frac{\mu_2\mu_5}{\mu_3}d\psi^2+k\frac{\mu_2\mu_5}{\mu_3}\frac{\prod_{i<j}(\rho^2+\mu_i\mu_j)\left[d\rho^2+dz^2\right]}{(\rho^2+\mu_1\mu_4)^3(\rho^2+\mu_2\mu_5)^3\prod_{i=1}^5(\rho^2+\mu_i^2)} \ , \label{doubletang}\eequ
where $k$ is an integration constant. One can verify that the metric
is invariant under the exchange
\bequ
(a_1,a_2)\leftrightarrow (a_4,a_5) \ , \label{invariancestatic}\eequ
which generalises (\ref{invarianceback}) in the presence of static black holes. Taking $k=1$ and the periodicities $\Delta \phi=2\pi=\Delta \psi$ guarantees this solution is asymptotically flat. There are, however, conical singularities for $a_2< z<a_3$ and $a_3<z< a_4$ in the $\rho-\psi$ and $\rho-\phi$ planes, respectively. The conical \textit{excesses} are, respectively \cite{Tan:2003jz}
\bequ
\barr{c}
\displaystyle{
\delta_{\psi}=2\pi\left(\frac{a_{41}a_{52}}{\sqrt{a_{51}a_{31}a_{32}a_{42}}}-1\right)
\ , \ \ a_2\le z<a_3 \ ;}\\ \\
\displaystyle{\delta_{\phi}=2\pi\left(\frac{a_{41}a_{52}}{\sqrt{a_{51}a_{43}a_{53}a_{42}}}-1\right) \ , \ \ a_3<z \le a_4 \ .} 
\label{excessesstatic} 
\earr
\eequ
It is straightforward to show that, for the ordering (\ref{ordering}), $\delta_{\psi}$ and $\delta_{\phi}$ are strictly positive. Hence \textit{there is no choice of parameters that makes the background free of conical singularities in the static case.} We will contrast this state of affairs with that of the stationary solution presented below.

The mass, area and temperature of each black hole can be written as 
(we set the five dimensional Newton constant to one)
\bequ
M_1^{Komar}=\frac{3\pi}{8}\,\Delta  \ , \ \ \ \ 
\mathcal{A}_{1}=2\pi^2\frac{\sqrt{2a_{21}a_{31}a_{51}}}%{|a_{41}|}\,\Delta  \ , \ \ \ \
{a_{41}}\,\Delta  \ , \ \ \ \
{T}_1=\frac{1}{2\pi}\sqrt{\frac{2a_{21}}{a_{31}a_{51}}}\,%\frac{|a_{41}|}{\Delta} \ ,
\frac{a_{41}}{\Delta} \ ,  
\label{bh1static} \eequ
\bequ
M_2^{Komar}=\frac{3\pi}{8}\,\bar{\Delta} \ , \ \ \ \ 
\mathcal{A}_{2}=2\pi^2\frac{\sqrt{2a_{54}a_{53}a_{51}}}%{|a_{52}|}\,\bar{\Delta} \ , \ \ \ \
{a_{52}}\,\bar{\Delta} \ , \ \ \ \
{T}_2=\frac{1}{2\pi}\sqrt{\frac{2a_{54}}{a_{53}a_{51}}}\,%\frac{|a_{52}|}{\bar{\Delta}} \ ,
\frac{a_{52}}{\bar{\Delta}} \ ,  
\label{bh2static} \eequ
where the individual masses can be computed as Komar integrals at each horizon 
(cf. section \ref{komar}), and for reasons that will become clear later we introduced
\bequ
\Delta\equiv 2a_{21} \ , \ \ \ \ \ \ \bar{\Delta}\equiv 2a_{54} \ . \label{deltas} \eequ
These quantities are consistent with the Smarr-type formula
\bequ
\frac{2}{3}\,M_i^{Komar}=T_i\,\frac{\mathcal{A}_i}{4} \ , \ \ \ \ i=1,2 \ . \eequ
Note also that in this case the two black hole masses add up to the ADM mass of the spacetime:
\bequ
M_{ADM}=M_1^{Komar}+M_2^{Komar} \ . \eequ
Finally, observe that under (\ref{invariancestatic}) the physical masses and 
conical excesses are interchanged, as one would expect:
\[
M_1^{Komar}\leftrightarrow M_2^{Komar}\ ,\ \ \ \ \ \ \ \ \ \ \ \
\delta_{\psi}\leftrightarrow \delta_{\phi} \ . 
\]

\section{The stationary case: double Myers-Perry}
\subsection{Generating the solution with the inverse scattering method} \label{sec:seed}
In $D$ spacetime dimensions, the inverse scattering method (or Belinskii-Zakharov method) \cite{Belinskii:78,Belinskii:79} can be used to construct new Ricci flat metrics with $D-2$ commuting Killing vector fields from known ones,  by using purely algebraic manipulations. Such metrics can always be written in the form
\bequ
ds^2=G_{ab}(\rho,z)dx^adx^b+e^{2\nu(\rho,z)}(d\rho^2+dz^2) \ , \eequ
where $a,b=1,\dots,D-2$.  In what follows we shall specialise all results of the method to the case of interest herein; in particular $D=5$.

The  seed metric is the double Schwarzschild-Tangherlini spacetime (\ref{doubletang}):
\begin{equation}
 G_0 = \rm{diag}  \left\{ - \frac{\mu_1 \mu_4}{\mu_2 \mu_5}, -\frac{\bar{\mu}_1 \mu_3}{\mu_4}, \frac{\mu_2 \mu_5}{\mu_3} \right\} \ . \label{g0}
\end{equation}
As usual the $\mu$'s refer to soliton positions in the BZ method and 
\[\tilde{\mu}_k=\pm\sqrt{\rho^2+(z-a_k)^2}-(z-a_k) \ ; \]
the ``$+$'' pole refers to a \textit{soliton} and is denoted by $\mu_k$; the ``$-$'' 
pole refers to an \textit{anti-soliton} and is denoted by $\bar{\mu}_k$. For the 
seed solution the conformal factor is, from (\ref{doubletang}), 
\bequ
e^{2\nu_0}=k\frac{\mu_2\mu_5}{\mu_3}\frac{\prod_{i<j}(\rho^2+\mu_i\mu_j)}{(\rho^2+\mu_1\mu_4)^3(\rho^2+\mu_2\mu_5)^3\prod_{i=1}^5(\rho^2+\mu_i^2)} \ , \label{enu}\eequ
where $k$ is an integration constant. 

We proceed with the method suggested by Pomeransky \cite{Pomeransky:2005sj} (see also \cite{Emparan:2008eg} for a recent review) and implement the following 4-soliton transformation:  we remove two anti-solitons, at $z=a_1$ and $z=a_4$, and two solitons, at $z=a_2$ and $z=a_5$,
 all with BZ vectors $(1,0,0)$. Thus we divide $(g_0)_{tt}$ by  $\rho^8/\bar{\mu}^2_1\bar{\mu}^2_4{\mu}^2_2{\mu}^2_5$. The seed metric becomes
\begin{equation}
 G_0'       = \frac{\mu_2 \mu_5}{\mu_1 \mu_4}\, {\rm{diag}}  \left\{ -1,\frac{\mu_3 \rho^2}{\mu_2\mu_5}, \frac{\mu_1 \mu_4}{\mu_3} \right\}        \equiv  \frac{\mu_2 \mu_5}{\mu_1 \mu_4} \tilde{G}_0 \ .
\end{equation}
We will actually take the rescaled metric $\tilde{G}_0$ to be our seed (bearing in mind that one should multiply the final metric by the overall factor $\mu_2 \mu_5/\mu_1 \mu_4 $).  We take the generating matrix to be 
\begin{equation}
 \tilde{\Psi}_0(\lambda,\rho,z)= {\rm{diag}}  \left\{-1,-\frac{(\bar{\mu_2}-\lambda)(\bar{\mu_5}-\lambda)}{(\bar{\mu_3}-\lambda)}, \frac{(\mu_1-\lambda) (\mu_4-\lambda)}{ (\mu_3-\lambda)} \right\}  \ .
\end{equation}
One can verify that this matrix solves the Lax pair constructed in the
BZ method (see \cite{Belinskii:78,Belinskii:79}).  The double
Myers-Perry solution is now obtained by a 4-soliton transformation:
using $\tilde{G}_0$ as seed, we add two anti-solitons, at $z=a_1$ with
BZ vector $m_{0b}^{(1)}=(1,b,0)$ and at $z=a_4$ with a BZ vector
$m_{0b}^{(4)}=(1,c,0)$, and add two trivial solitons, at $z=a_2$ with
BZ vector $m_{0b}^{(2)}=(1,0,0)$ and at $z=a_5$ with BZ vector
$m_{0b}^{(5)}=(1,0,0)$.\footnote{This 4-soliton transformation allows
  us to work with the simplest possible seed (rescaled metric
  $\tilde{G}_{0}$). Moreover, it would now be straightforward, even if
  computationally challenging, to generate the
  general double doubly spinning Myers Perry (i.e the solution where
  each  black hole has angular momentum in both planes)  just by considering  the non trivial BZ vectors: $m_{0b}^{(1)}=(1,b,0)$, $m_{0b}^{(2)}=(1,0,d)$, $m_{0b}^{(4)}=(1,c,0)$ and $m_{0b}^{(5)}=(1,0,e)$.} Notice that we have introduced two new parameters: $b$ and $c$.
 The resulting metric is 
\[G=\frac{\mu_2 \mu_5}{\mu_1 \mu_4} \tilde{G} \ , \]
where $\tilde{G}$ has components
%%%%%%%%%%%%
% \bequ
% \tilde{G}_{ab}=(\tilde{G}_0)_{ab}-\sum_{k,l=1}^4\frac{(\tilde{G}_0)_{ac} m_c^{(k)}\left(\tilde{\Gamma}^{-1}\right)_{kl} m_d^{(l)}(\tilde{G}_0)_{db}}{\tilde{\mu}_{k'}\tilde{\mu}_{l'}} \ ,\eequ
% where the primed index equals $(1,2,4,5)$ when the same unprimed index equals $(1,2,3,4)$, respectively, and $\tilde{\mu}_{k'}=\mu_{k'}$ for $k'=2,4$ whereas $\tilde{\mu}_{k'}=\bar{\mu}_{k'}$ for $k'=1,3$. The space-time components of the four vectors $m^{(k)}$ are given 
% by
% \bequ
% m^{(k)}_a=m_{0b}^{(k)}\left[\tilde{\Psi}_0^{-1}(\tilde{\mu}_{k'},\rho,z)\right]_{ba} \ , 
% \eequ
% and the four BZ vectors are taken to be
% \bequ
% m_{0b}^{(1)}=(1,b,0) \ , \ \ \ \ \ m_{0b}^{(2)}=(1,0,0) \ , \ \ \ \ \ 
% m_{0b}^{(3)}=(1,c,0) \ , \ \ \ \ \ m_{0b}^{(4)}=(1,0,0) \ . 
% \eequ
%%%%%%%%%%%%
\bequ
\tilde{G}_{ab}=(\tilde{G}_0)_{ab}-\sum_{k,l} \frac{(\tilde{G}_0)_{ac} m_c^{(k)}\left(\tilde{\Gamma}^{-1}\right)_{kl} m_d^{(l)}(\tilde{G}_0)_{db}}{\tilde{\mu}_{k}\tilde{\mu}_{l}} \ ,\eequ
with $k,l=1,2,4,5$ and $\tilde{\mu}_{k}=\mu_{k}$ for $k=2,4$ whereas $\tilde{\mu}_{k}=\bar{\mu}_{k}$ for $k=1,3$. The space-time components of the four vectors $m^{(k)}$ are given by
\bequ
m^{(k)}_a=m_{0b}^{(k)}\left[\tilde{\Psi}_0^{-1}(\tilde{\mu}_{k},\rho,z)\right]_{ba} \ . 
\eequ
The symmetric matrix $\tilde{\Gamma}$, whose inverse is $\tilde{\Gamma}^{-1}$, reads
%%%%%%
% \bequ
% \tilde{\Gamma}_{kl}=
% \frac{m_a^{(k)}(\tilde{G}_0)_{ab}m_b^{(l)}}{\rho^2+\tilde{\mu}_{k'}\tilde{\mu}_{l'}} \ . 
% \label{gamma} 
% \eequ
%%%%%%%
\bequ
\tilde{\Gamma}_{kl}=
\frac{m_a^{(k)}(\tilde{G}_0)_{ab}m_b^{(l)}}{\rho^2+\tilde{\mu}_{k}\tilde{\mu}_{l}} \ . 
\label{gamma} 
\eequ
Finally, it only remains to compute the function $\nu$ in the metric, which is given by
\bequ
e^{2\nu}=e^{2\nu_0}\frac{\det{\Gamma_{kl}}}{\det{\Gamma^{(0)}_{kl}}} \ , 
\eequ
where $\Gamma^{(0)}$ and $\Gamma$ are constructed as in (\ref{gamma}) using $G_0$ and $G$, respectively.

The end result of the above algorithm can be written in the following form, analogous to the black saturn solution \cite{Elvang:2007rd}
\begin{equation}
 ds^2 =  - \frac{H_y}{H_x} \left[ dt + \left(\frac{\omega_{\phi}}{H_y} - q \right) d \phi \right]^2
      + \frac{H_x}{H_y} \frac{ \rho^2 \mu_3}{\mu_2 \mu_5} d \phi^2
      + \frac{\mu_2 \mu_5}{\mu_3} d \psi^2 + k \frac{H_x}{F} (d\rho^2 + dz^2)\ ,\label{solution} \end{equation}
where a coordinate transformation $dt \to dt - q \, d\phi$ was performed; $q$ will be chosen below. The metric functions are\footnote{Following standard notation, the square roots of the function $M_i$ are to be understood as, for example, $\sqrt{\left(\mu _1-\mu _4\right)^2}=\mu _1-\mu _4$.}
\bequ
\barr{rl}
 H_x &= M_0 + b^2 M_1 + c^2 M_2 + bc M_3 + b^2c^2 M_4\ , \\ \\
 H_y &= \displaystyle{\frac{\rho^2}{\mu_2\mu_5}\left[M_0 \frac{\mu_1 \mu_4}{\rho^2} - b^2 M_1 \frac{\mu_4}{\mu_1} - c^2 M_2 \frac{\mu_1}{\mu_4} -bc M_3 
 + b^2c^2M_4 \frac{\rho^2}{\mu_1 \mu_4 }\right]}\ , \\ \\
 \omega_\phi & = \displaystyle{2\sqrt{\frac{\mu_3}{\mu_2\mu_5}}\left[b R_1\sqrt{M_0M_1} + c R_4 \sqrt{M_0M_2} - b^2 c R_4\sqrt{M_1M_4} - b c^2 R_1\sqrt{M_2M_4} \right]} \ ;
\earr
\eequ
where $R_i=\sqrt{\rho^2+(z-a_i)^2}$ and the functions $M_i$ are
\bequ
\barr{rl}
 M_0 &\equiv \mu _2\mu_3^2\mu _5  \left(\mu _1-\mu
 _4\right)^2\left(\rho ^2+\mu _1 \mu _2\right)^2 \left(\rho ^2+\mu _1
 \mu _5\right)^2 \left(\rho ^2+\mu_2 \mu _4\right)^2 \left(\rho ^2+\mu
 _4 \mu_5\right)^2,  \\ \\
 M_1 &\equiv \mu_1^2\mu_2^2 \mu_3\mu_5^2 \left(\mu _1-\mu _3\right)^2
 \left(\rho ^2+\mu _1\mu_4 \right)^2\left(\rho ^2+\mu _2 \mu
 _4\right)^2  \left(\rho ^2+\mu _4 \mu _5\right)^2\ , \\ \\
 M_2 &\equiv \mu_2^2 \mu_3\mu_4^2\mu_5^2 \left(\mu _3-\mu
 _4\right)^2\left(\rho ^2+\mu _1 \mu _2\right)^2  \left(\rho ^2+\mu _1
 \mu _4\right)^2  \left(\rho ^2+\mu _1 \mu _5\right)^2\ ,  \\ \\
 M_3 &\equiv 2 \mu _1\mu _2^2\mu _3\mu _4\mu _5^2\left(\mu _1-\mu _3\right)\left(\mu _3-\mu _4\right) \left(\rho ^2+\mu _1^2\right)\left(\rho ^2+\mu _4^2\right) \left(\rho^2+\mu _1 \mu _2\right)\left(\rho ^2+\mu _1 \mu _5\right)    \\
   & ~~~\times 
      \left(\rho ^2+\mu _2 \mu _4\right)\left(\rho ^2+\mu _4\mu
      _5\right), \\ \\
 M_4 &\equiv \mu _1^2 \mu _2^3\mu _4^2 \mu _5^3\rho ^4  \left(\mu _1-\mu _3\right)^2  \left(\mu _1-\mu _4\right)^2\left(\mu _3-\mu _4\right)^2 \ .
\earr
\eequ
Moreover
\bequ
\barr{rl}
F & =
 \mu _3^3  \left(\mu _1-\mu _4\right)^2 \left(\rho^2+\mu _1 \mu _2\right) \left(\rho^2+\mu _1 \mu _4\right)^2 \left(\rho ^2+\mu _1 \mu _5\right)\left(\rho ^2+\mu _2 \mu _5\right)^2 \left(\rho ^2+\mu _2 \mu _4\right) \nonumber \\
  &~~~\times
    \left(\rho ^2+\mu _4
   \mu _5\right)\prod_{i=1}^5 \left(\rho ^2+\mu _i^2\right)/\left[\left(\rho ^2+\mu _1 \mu _3\right) \left(\rho ^2+\mu _2 \mu _3\right)
   \left(\rho ^2+\mu _3 \mu _4\right)\left(\rho ^2+\mu _3 \mu
 _5\right)\right] \ . \label{F}
\earr
\eequ

The metric (\ref{solution}) is invariant under the exchange 
\bequ (a_1,a_2,b)\leftrightarrow (a_4,a_5,c) \ , 
\label{invariance}
\eequ
which generalises (\ref{invarianceback}) and (\ref{invariancestatic}) 
to the case of two stationary black holes.

Let us note that, despite the high degree of complexity of this solution, 
it is drastically simpler than the four dimensional double Kerr solution, 
originally obtained via a B\"acklund transformation \cite{Kramer:79}. From 
the viewpoint of the inverse scattering method this can be understood from 
the fact that the double Kerr can only be constructed, from a double 
Schwarzschild seed, by a four-soliton transformation with \textit{all} 
solitons having non-trivial BZ vectors.

\subsection{Rod structure, horizons angular velocities and axis condition}
\label{axissection}
\begin{figure}[h!]
\begin{picture}(0,0)(0,0)
\put(84,74){$_{(1,\Omega^{\phi}_1,0)}$}
\put(198,74){$_{(1,\Omega^{\phi}_2,0)}$}
\put(30,43){$_{(0,1,0)}$}
\put(163,43){$_{(h,1,0)}$}
\put(260,12){$_{(0,0,1)}$}
\put(126,12){$_{(0,0,1)}$}
\put(-2,62){$t$}
\put(-5,33){$\phi$}
\put(-5,2){$\psi$}
\put(73,-5){$a_1$}
\put(111,-5){$a_2$}
\put(151,-5){$a_3$}
\put(190,-5){$a_4$}
\put(228,-5){$a_5$}
\end{picture}
\centering\epsfig{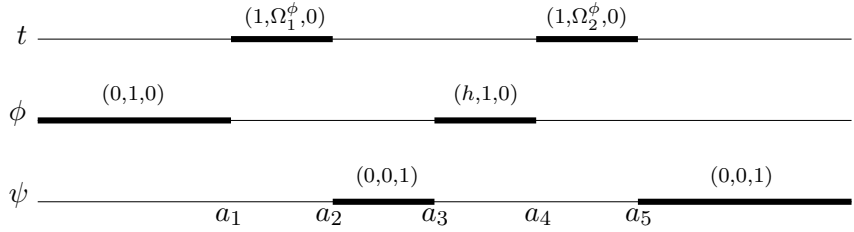}
\caption{Rod structure for the double Myers-Perry spacetime. Next to each rod 
the corresponding eigenvector \cite{Harmark:2004rm} is displayed.}
\label{doublemp}
\end{figure}

The rod structure of the solution we have just generated is the same as the 
one of the static solution, except for the directions of the 
rods - figure \ref{doublemp}. From this rod structure it is clear that
the metric gives a six parameter family of solutions. The parameters can be
taken to be the four finite rod sizes, together with $b$ and
$c$. Physically, the six independent degrees of freedom can be taken
to be the two black hole masses and angular momenta, together with the
two conical singularities. Alternatively one can replace the two conical
singularities by the two distances $d_1\equiv a_{32}$ and
$d_2\equiv a_{43}$. Note that $d=d_1+d_2$
is the (coordinate) distance squared between the two black holes. 

%\spa{0.05cm}
\bigskip

The eigenvector of the two timelike rods gains a spatial component, along the 
$\phi$ direction. These new components are the angular velocities of the 
individual black hole horizons. A computation shows that they take the form
\bequ
\Omega^{\phi}_1=\frac{a_{41}b}{a_{51}\Delta} \ , \ \ \ \ \ \ \ \ \ 
\Omega^{\phi}_2=\frac{a_{54}\tilde{b}+a_{51}\tilde{c}}{a_{41}a_{51}\bar{\Delta}} \ ,  
\eequ
where, for convenience, we have introduced the quantities
\bequ \Delta\equiv 2a_{21}+\frac{a_{31}}{a_{51}}b^2\ , \ \ \ \   
\bar{\Delta}\equiv 2a_{54}+
\frac{(\tilde{b}+\tilde{c})(a_{54}\tilde{b}+a_{51}\tilde{c})}{a_{41}^2a_{51}}  \ ,  
\eequ 
which generalise (\ref{deltas}) for the stationary case, and  
\bequ \tilde{b}\equiv a_{31}b \ ,  \ \ \ \ \tilde{c}  \equiv a_{43}c \ . 
\eequ
These angular velocities reduce to the horizon angular velocities of single 
Myers-Perry black holes  in the limits $a_3=a_4=a_5$ and $a_1=a_2=a_3$, 
respectively (cf. (\ref{bh1}) and (\ref{bh2})). 

\begin{figure}[h!]
\begin{picture}(0,0)(0,0)
\put(-8,200){$\tilde{b}\tilde{c}$}
\put(-40,181){$_{2a_{51}a_{42}a_{43}}$}
\put(-4,102){$0$}
\put(-72,80){$_{-2a_{51}(a_{41}^2-a_{42}a_{43})}$}
\put(130,95){$_{2a_{51}a_{42}a_{43}} $}
\put(310,93){$\tilde{c}^2$}
\end{picture}
\centering\epsfig{file=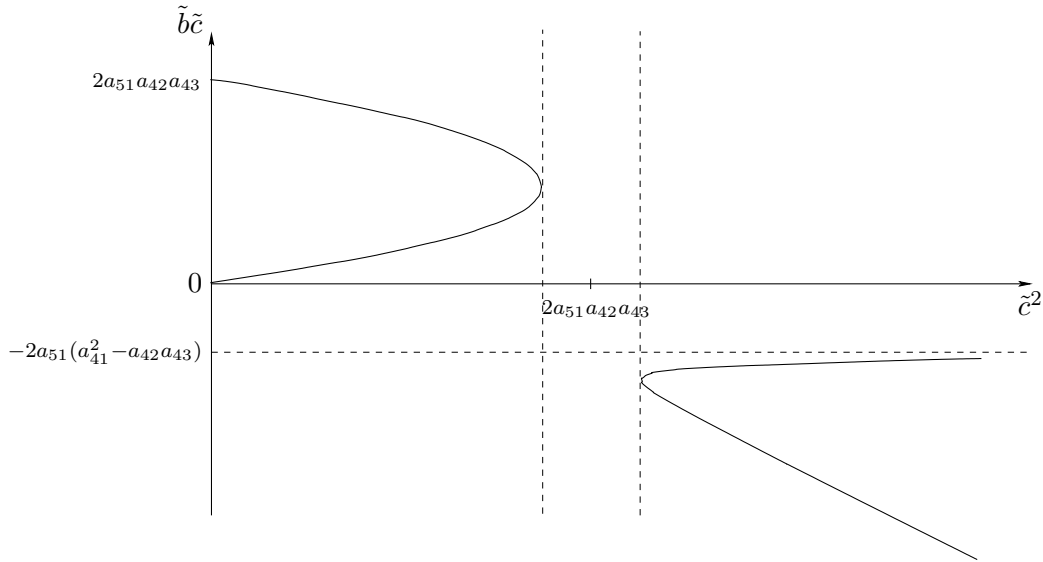,width=11cm}
\caption{Axis condition (4.18): $\tilde{b}\tilde{c}$
  can only take the values (4.20).}
\label{bc}
\end{figure}

The finite rod between $a_3$ and $a_4$ (figure \ref{doublemp}) 
also gains a timelike component, 
\[
h=-\left(\frac{g_{\phi\phi}}{g_{t\phi}}\right)_{\rho=0,\, a_3<z<a_4}= \frac{(\tilde{b}+\tilde{c})(2a_{42}a_{51}-\tilde{b}c)-2a^2_{41}a_{51}c 
 }{a_{41}(2a_{42}a_{51}-\tilde{b}c)} \ . 
\]  
Thus $h= 0$ iff $(g_{\phi\phi})_{\rho=0,\, a_3<z<a_4}=0$. The latter is 
sometimes called the \textit{axis condition} \cite{Bonnor:2001} 
(see also \cite{Letelier:1998ft}); if violated, $\rho=0$ and $a_3<z<a_4$ is 
not an axis for $\partial/\partial \phi$; moreover, if $h\neq0$ there are naked 
closed timelike curves in spacetime for some choices of $b$ and $c$, 
which are generically regarded as pathological. Thus, we demand $h=0$, which 
yields the constraint
\bequ
\Delta_{axis}=0\ ,\ \ \ \ \ \ \ 
\Delta_{axis}\equiv (\tilde{b}+\tilde{c})(2a_{42}a_{51}-\tilde{b}c)-2a^2_{41}a_{51}c \ . 
\label{axiscondition} 
\eequ
In particular, this equation is obeyed if $b=0=c$, as expected. It is also obeyed if we take the limit in which the first black hole disappears, i.e $a_1=a_2=a_3$. Note that
it does not make sense to consider the limit of (\ref{axiscondition}) in which the second black hole disappears, i.e $a_3=a_4=a_5$, since in that limit the rod whose direction defines the axis condition collapses to zero size. In general, (\ref{axiscondition}) can be regarded as an
equation defining $\tilde{c}^2$ in terms of $\tilde{b}\tilde{c}$:
\bequ
\tilde{c}^2=
\frac{(2a_{42}a_{43}a_{51}-\tilde{b}\tilde{c})\,\tilde{b}\tilde{c}}
{\tilde{b}\tilde{c}+2a_{51}(a_{41}^2-a_{42}a_{43})} \ .
\label{csquare}
\eequ
Positivity of the left hand side restricts the possible values of
$\tilde{b}\tilde{c}$ to
\bequ
-\infty <\tilde{b}\tilde{c}<-2a_{51}(a_{41}^2-a_{42}a_{43})\ \ \     \vee \ \ \ 0<\tilde{b}\tilde{c}<2a_{51}a_{42}a_{43} \ , \label{bcconstraint}\eequ
as displayed in figure \ref{bc}.

\section{Analysis of the double Myers-Perry solution}

\subsection{Single black hole limits}
Let us now see that the solution (\ref{solution}) indeed contains two 
Myers-Perry black holes. First collapse the rod structure of the second 
black hole by taking $a_3=a_4=a_5$. To establish that the resulting metric 
describes a Myers-Perry black hole with a single angular momentum parameter 
it is convenient to change from Weyl canonical coordinates $(\rho,z)$ to 
prolate spherical coordinates $(x,y)$ by
\[ 
\mu_{1,2}=\alpha(x\mp 1)(1-y) \ , \ \ \ \ 2\alpha\equiv a_{21} \ , 
\]
so that $\rho^2=\alpha^2(x^2-1)(1-y^2)$. Defining also $\rho_0^2\equiv 4\alpha+b^2$, 
the metric coefficients become (take $q=b$ so that $g_{t\phi}\rightarrow 0$ 
asymptotically)
\bequ
G_{tt}=-\frac{4\alpha x- b^2y-\rho_0^2}{4\alpha x- b^2y+\rho_0^2} \ , 
\ \ \ \ \ \ \ \ 
G_{t\phi}=-\frac{b\rho_0^2(1+y)}{4\alpha x-b^2y+\rho_0^2} \ ,
\nonumber 
\eequ
\bequ
G_{\phi\phi}=\frac{1+y}{4}
\left(4\alpha x+b^2+\rho_0^2+\frac{2b^2\rho_0^2(1+y)}{4\alpha x -b^2y+\rho_0^2}\right) \ , 
\nonumber 
\eequ
\bequ
G_{\psi\psi}=\alpha(1-y)(1+x) \ , \ \ \ \ \ \ \ 
e^{2\nu}=k\frac{H_x}{F}=\frac{4\alpha x-b^2y+\rho_0^2}{8\alpha^2(x^2-y^2)} \ , 
\nonumber
\eequ
where we have taken $k=1$ and a standard $2\pi$ period for the azimuthal angles; 
this choices make the geometry free of conical singularities. The above metric 
coefficients coincide with those of the Myers-Perry black hole with one angular 
momentum \cite{Harmark:2004rm}. The ADM mass (which equals the Komar mass), 
ADM angular momentum (which equals the Komar angular momentum), horizon angular 
velocity, area and temperature of this black hole are given, respectively, by
\bequ 
M_1^{Komar}=\frac{3\pi}{8}\,\Delta_1 \ , \ \ \ \  
J_1^{\phi}=\frac{\pi}{4}\,b\,\Delta_1  \ , \ \ \ \ 
\Omega_1^{\phi}=\frac{b}{\Delta_1}\ , \ \ \ \ 
\mathcal{A}_1=2\pi^2\sqrt{2a_{21}}\,\Delta_1 \ , \ \ \ 
T_1=\frac{1}{2\pi}\frac{\sqrt{2a_{21}}}{\Delta_1} \ , 
\label{bh1}
\eequ
where
\[ \Delta_1\equiv 2a_{21}+b^2 \ . \]

Similarly we can collapse the rod structure of the first black hole by 
taking $a_1=a_2=a_3$. All of the above steps can be repeated, with the 
replacements $a_{21}\rightarrow a_{54}$ and $b\rightarrow c$. One finds 
another Myers-Perry black hole, with ADM mass, ADM angular momentum, 
horizon angular velocity, area and temperature given by
\bequ 
M_2^{Komar}=\frac{3\pi}{8}\,\Delta_2 \ , \ \ \ \  
J_2^{\phi}=\frac{\pi}{4}\,c\,\Delta_2 \ , \ \ \ \ 
\Omega_2^{\phi}=\frac{c}{\Delta_2} \ ,\ \ \ \ 
\mathcal{A}_2=2\pi^2\sqrt{2a_{54}}\,\Delta_2 \ , \ \ \ 
T_2=\frac{1}{2\pi}\frac{\sqrt{2a_{54}}}{\Delta_2} \ , \label{bh2}\eequ
where
\[ \Delta_2\equiv 2a_{54}+c^2 \ . \]
Note that the extremal limit of black hole 1 (black hole 2) is obtained as $a_{21}\rightarrow 0$ ($a_{54}\rightarrow 0$), for $b\neq 0$ ($c\neq 0$). Note also that each of these black holes obeys a Smarr-type formula:
\bequ
\frac{2}{3}\,M_i^{Komar}=T_i\,\frac{\mathcal{A}_i}{4}+\Omega^{\phi}_iJ_i^{\phi}\ , \ \ \ \ 
i=1,2 \ . 
\label{smarrgeneral} 
\eequ

\subsection{Asymptotics and  physical quantities}
We now show that the solution is asymptotically flat and read off the
ADM mass and angular momentum. Introducing the asymptotic coordinates $r$ and $\theta$
\bequ
  \rho = \frac{1}{2} r^2 \sin{2\theta} \ , \qquad
  z = \frac{1}{2} r^2 \cos{2 \theta}\ ,
\eequ
the asymptotic limit becomes $r \to \infty$. We can check that $G_{tt} = -1 + O\left(\frac{1}{r^2} \right) $, as expected, and fix $q$ by requiring that $G_{t \phi} \to 0$ as $r \to \infty $; this yields
\begin{equation}
  q=\frac{\tilde{b} + \tilde{c}}{a_{41}}\ .
\end{equation}
For the conformal factor $e^{2 \nu(\rho,z)}$ we have, asymptotically,
\begin{equation}
  e^{2\nu} = \frac{k}{r^2} + O\left(\frac{1}{r^4}\right) \ ,
\end{equation}
which fixes $k=1$. Thus, at infinity, the metric reduces to the standard form in bipolar coordinates
\begin{equation}
  ds^2 = -dt^2 + dr^2 + r^2 d\theta^2 + r^2 \sin^2{\theta} d \psi^2 + r^2 \cos^2{\theta} d\phi^2\ ,
\end{equation}
so that taking the canonical periods $ \Delta \phi =2\pi= \Delta \psi$ guarantees absence of conical singularities at infinity.

From the  next to leading order term in $G_{tt}$ and leading order term in $G_{t\phi}$ we can read off the ADM mass and angular momentum to be 
\bequ
M_{ADM}=
\frac{3\pi}{8}\left[2a_{21}+2a_{54}+\frac{(\tilde{b}+\tilde{c})^2}{a_{41}^2}\right] \ , 
\eequ
\bequ 
J^{\phi}_{ADM}=\frac{\pi}{4}\left[\frac{2[\tilde{b}(a_{21}+a_{54}+a_{34})+\tilde{c}(a_{21}+a_{54}+a_{31})]}{a_{41}}+\frac{(\tilde{b}+\tilde{c})^3}{a_{41}^3}\right] \ . 
\eequ
Note that these expressions i) are invariant under (\ref{invariance}) as one would expect; ii) reduce to (\ref{bh1}) and (\ref{bh2}) in the limits $a_3=a_4=a_5$ and $a_1=a_2=a_3$, respectively. Note also that the ordering (\ref{ordering}) guarantees positivity of the ADM mass.

\subsection{Conical singularities}
\label{conicalsingularitiessec}
The conical \textit{excesses} for the generic solution are 
\bequ
\delta_{\psi}=2\pi\left(\frac{a_{41}a_{52}}{\sqrt{a_{51}a_{31}a_{32}a_{42}}}\cdot
\left|\frac{2a_{42}a_{51}}{2a_{42}a_{51}+b\tilde{c}}\right|-1\right) \ , \ \ \ \ \ 
a_2\le z<a_3 \ ; 
\label{deltapsi} 
\eequ 
\bequ 
\delta_{\phi}=2\pi\left(\frac{a_{41}a_{52}}{\sqrt{a_{51}a_{43}a_{53}a_{42}}}\cdot
\left|\frac{2a_{42}a_{51}}{2a_{42}a_{51}-\tilde{b}c}\right|-1\right) \ , \ \ \ \ \ 
a_3<z\le a_4 \ . 
\label{excessesstationary} 
\eequ
These reduce to (\ref{excessesstatic}) when $b=0=c$ and to (\ref{cexy}) 
if also $a_1=a_2$ and $a_4=a_5$. Note that the second condition should only be considered if one imposes the axis condition.

It is clear that the introduction of rotation could eliminate either of 
these conical singularities, but not \textit{both}
simultaneously. However, one must note that the requirement for either
of these conical singularities to vanish is \textit{incompatible} with
the axis condition. To see this, require first $\delta_{\psi}=0$. 
This demands $bc>0$. We already know that
the axis condition puts an upper bound on the positive values of $bc$;
thus, we can parametrise the possible values of $bc$ as 
\[ bc=\frac{2a_{51}a_{42}}{a_{31}}\epsilon \ , \ \ \ \ \ \ \ \ 
0\le \epsilon \le 1 \ . 
\]
Substituting in (\ref{excessesstationary}) we observe that $\epsilon\neq 1$ 
do avoid a divergence in $\delta_{\phi}$. The condition that $\delta_{\psi}=0$ 
becomes
\[ 
\epsilon=\frac{a_{31}}{a_{43}}
\left(\frac{a_{41}a_{52}}{\sqrt{a_{51}a_{31}a_{32}a_{42}}}-1\right) \ . 
\]
It is fairly simple to show that the RHS of this last equation is
always greater or equal to 1; since we have seen that the LHS is
smaller than one we can conclude that \textit{$\delta_{\psi}$ cannot
  be set to zero and, at the same time, obey  the axis condition.} Thus we can set $\delta_{\psi}=0$, which regularises this conical singularity but, generically, the geometry will develop closed timelike curves.

Let us now  require $\delta_{\phi}=0$. This demands $bc<0$, in fact 
\bequ bc=\frac{2a_{42}a_{51}}{a_{31}}(1-\beta) \ , \ \ \ \ \ \beta\equiv
\frac{a_{41}a_{52}}{\sqrt{a_{51}a_{43}a_{53}a_{42}}} \ . \label{bcregular}\eequ
Note that $\beta\ge 1$. Replacing in (\ref{csquare}) one gets
\[
c^2=\frac{2a_{42}^2a_{51}\beta(1-\beta)}{a_{41}^2-a_{42}a_{43}\beta} \ ,
\]
whose RHS is manifestly negative (observe that $a_{41}^2\ge a_{42}a_{43}\beta$). Thus \textit{$\delta_{\phi}$ cannot be set to zero and, at the same time, obey  the axis condition.} Notice therefore that $\delta_{\phi}=0$ cannot be interpreted as a regularity condition, since, when it is obeyed, $\rho=0$, $a_3<z<a_4$ is not an axis.

The incompatibility of the axis and regularity conditions is
reminiscent of the result obtained in \cite{Bonnor:2001} for $D=4$ using a 
post-post Newtonian analysis.

\subsection{Horizons geometry, areas and temperatures}
Let us now show that both black holes have, in general, regular
(except for a conical singularity at one point) finite area horizons
and finite temperatures.

The horizon of the first black hole is located at $\rho=0$ and 
$ a_1 < z < a_2 $. Considering the coordinate transformation
\bequ
\rho=\frac{1}{2}\sqrt{1-\frac{2 a_{21}} {R^2}} \,R^2 \sin{2 \theta}\ , \ \ \ \ \ 
z=\frac{a_1 + a_2}{2} + \frac{1}{2} \left(R^2 - a_{21}\right)\cos{2\theta} \ ,
\eequ
the horizon is located at $ R^2=2a_{21}$. Note that 
$z = a_2 - a_{21}\sin^2{\theta} $. 
The metric on a spatial section of the horizon reads
\bequ
\barr{l}
\displaystyle{ds_{H_1}^2=  
\frac{a_{31}a_{51}}{a^2_{41}}\,\Sigma(\theta)f_1(\theta)\,d\theta^2}+
\frac{f_2(\theta)\cos^2\theta}{\Sigma(\theta)}\,\Delta^2\,d\phi^2+
2a_{21}f_3(\theta)\sin^2\theta \,d\psi^2  \ , 
\earr 
\label{bhhorizon1}
\eequ 
where 
\[ 
\Sigma(\theta)\equiv F_1(\theta)\sin^2\theta+2a_{21}\big(1+F_2(\theta)\big)^2 \ ,
\]
with the functions $F(\theta)$ given by
\[
F_1(\theta)\equiv 
b^2\frac{a_{41}^2}{a_{51}^2}\frac{f_2(\theta)}{f_1(\theta)} \ , \ \ \ \ 
F_2(\theta)\equiv 
bc \frac{a_{43}}{a_{51}}\frac{\cos^2\theta}{2(a_{42}+a_{21}\sin^2\theta)} \ , 
\]
and the functions $f(\theta)$ given by
\bequ
 f_1(\theta)  \equiv\frac{ a_{42} + a_{21} \sin^2 {\theta} } {a_{52} + a_{21} \sin^2 {\theta}}\ , \ \ \ \ \ 
 f_2(\theta)  \equiv \frac{ a_{32} + a_{21} \sin^2 {\theta} } {a_{42} + a_{21} \sin^2 {\theta}}\ , \ \ \ \ \
 f_3(\theta)  \equiv \frac{ a_{52} + a_{21} \sin^2 {\theta} } {a_{32} + a_{21} \sin^2 {\theta}}\ . \label{f} \eequ
The area and temperature of this black hole are given by
\bequ
\mathcal{A}_{1}=2\pi^2\frac{\sqrt{2a_{21}a_{31}a_{51}}}{a_{41}}\,\Delta \ , \ \ \ \
{T}_1=\frac{1}{2\pi}\sqrt{\frac{2a_{21}}{a_{31}a_{51}}}\,\frac{a_{41}}{\Delta} \ . 
\label{atbh1} 
\eequ
Note that the results (\ref{bhhorizon1})-(\ref{atbh1}) reduce to the
expressions in \cite{Tan:2003jz}, for $b=0$, and to the ones of a single Myers-Perry black hole  for $a_3=a_4=a_5$, in particular to (\ref{bh1}).

A similar analysis can be done for 
the horizon of the second black hole, which is located at $\rho=0$ and 
$ a_4 < z < a_5 $. Considering the coordinate transformation
\bequ
\rho=\frac{1}{2} \sqrt{1 - \frac{2 a_{54}} {R^2}} \,R^2 \sin{2 \theta}\ , \ \ \ \ \ 
z= \frac{a_4 + a_5}{2} + \frac{1}{2} \left( R^2 - a_{54} \right) \cos{2 \theta} \ ,
\eequ
the horizon is located at $ R^2=2a_{54}$. Note that $z = a_5 - a_{54}
\sin^2{\theta} $. The metric on a spatial section of the horizon reads
\bequ
\barr{l}
\displaystyle{ds_{H_2}^2=  
\frac{a_{53}a_{51}}{a^2_{52}}\,\bar{\Sigma}(\theta)\bar{f}_1(\theta)\,d\theta^2 +
\frac{\bar{f}_2(\theta)\cos^2\theta}{\bar{\Sigma}(\theta)}\,\bar{\Delta}^2\,d\phi^2+
2a_{54}\bar{f}_3(\theta)\sin^2\theta \,d\psi^2  } \ . 
\earr 
\label{bhhorizon2}
\eequ
where 
\[ \bar{\Sigma}(\theta)\equiv \bar{F}_1(\theta)\sin^2\theta + 2a_{54}(1+\bar{F}_2(\theta))\ ,  \]
with the functions $\bar{F}(\theta)$ given by
\[
\bar{F}_1(\theta)\equiv \frac{\tilde{c}^2}{a_{41}^2}\frac{\bar{f}_2(\theta)}{\bar{f}_1(\theta)} \ , \ \ \ \  \bar{F}_2(\theta)=\frac{\tilde{b}\, a_{54}\sin^2\theta\cos^2\theta}{a^2_{41}a_{51}^2(a_{42}+a_{54}\cos^2\theta)}\left[a_{51}\tilde{c}\bar{f}_2(\theta)+\frac{\tilde{b}\, a^2_{54}\cos^2\theta}{2(a_{43}+a_{54}\cos^2\theta)}\right] \ , \]
and the functions $\bar{f}(\theta)$ given by
\bequ
 \bar{f}_1(\theta)  \equiv\frac{ a_{42} + a_{54} \cos^2 {\theta} } {a_{41} + a_{54} \cos^2 {\theta}}\ , \ \ \ \ \ 
 \bar{f}_2(\theta)  \equiv \frac{ a_{41} + a_{54} \cos^2 {\theta} } {a_{43} + a_{54} \cos^2 {\theta}}\ , \ \ \ \ \
 \bar{f}_3(\theta)  \equiv \frac{ a_{43} + a_{54} \cos^2 {\theta} } {a_{42} + a_{54} \cos^2 {\theta}}\ . \label{barf}\eequ
The area and temperature of this black hole are given by
\bequ
\mathcal{A}_{2}=2\pi^2\frac{\sqrt{2a_{54}a_{53}a_{51}}}{a_{52}}\,\bar{\Delta} \ , \ \ \ \
{T}_2=\frac{1}{2\pi}\sqrt{\frac{2a_{54}}{a_{53}a_{51}}}\,\frac{a_{52}}{\bar{\Delta}} \ . 
\label{atbh2} 
\eequ
Note that the results (\ref{bhhorizon2})-(\ref{atbh2}) reduce to the expressions 
in \cite{Tan:2003jz}, for $b=0=c$, and to the ones of a single Myers-Perry black 
hole  for $a_1=a_2=a_3$, in particular to (\ref{bh2}).

The surfaces described by (\ref{bhhorizon1}) and (\ref{bhhorizon2})
are topologically 3-spheres; they are regular, for generic parameters,
except for a conical singularity at $\theta=0$ for the first black hole,  
where there is a conical excess in $\psi$ given by (\ref{deltapsi}), and at
$\theta=\pi/2$  for the second black hole, where there is a conical excess in
$\phi$ given by (\ref{excessesstationary}), \textit{if one imposes}
the axis condition (\ref{axiscondition}).  Note that at $\theta=\pi/2$ for the first black hole, and at $\theta=0$ for the second, 
there are no conical singularities, in agreement with our discussion of
section \ref{background}.

\subsection{Individual masses and angular momenta}
\label{komar}
The individual mass of each black hole can be computed as a Komar integral at the horizon of each black hole. In five dimensions, and for a metric of type (\ref{solution}) the integral takes the form
\[
M^{Komar}=\frac{3}{32\pi G_5}\int_{S}\star d\xi=\frac{3}{32\pi G_5}
\int_{H_i} dz d\phi d\psi \,\frac{g_{\rho\rho}g_{\psi\psi}}{\sqrt{-g}}
\left[g_{t\phi}g_{t\phi,\rho}-g_{\phi\phi}g_{tt,\rho}\right] \ , \]
where $\xi=g_{tt}dt+g_{t\phi}d\phi$ is the one-form dual to the
asymptotic time translations Killing vector field
$\partial/\partial_t$ and $S$ is the boundary of any spacelike
hypersurface; to derive the second equality we have already chosen $S$ to
be a spatial section of the event horizon of one of the two black
holes. Thus $a_1<z<a_2$ ($a_4<z<a_5$)  for the first (second) black
hole. We find
\bequ
M^{Komar}_1=\frac{3\pi}{8}\frac{2a_{42}a_{51}}{2a_{42}a_{51}+b\tilde{c}}\,\Delta
\ , \ \ \ \ \ \
M^{Komar}_2=\frac{3\pi}{8}\,\bar{\Delta} \ . \label{komarmass}\eequ

The intrinsic spin of each black hole can also be computed as a Komar integral at the horizon of each black hole. In five dimensions, and for a metric of type (\ref{solution}) the integral takes the form
\bequ
J^{Komar}=-\frac{1}{16\pi G_5}\int_{S}\star d\zeta=
-\frac{1}{16\pi G_5}\int_{H_i} dz d\phi d\psi\,\frac{g_{\rho\rho}g_{\psi\psi}}{\sqrt{-g}}
\left[g_{t\phi}g_{\phi\phi,\rho}-g_{\phi\phi}g_{t\phi,\rho}\right] \ , 
\label{komarj} 
\eequ
where $\zeta=g_{\phi\phi}d\phi+g_{t\phi}dt$ is the one-form dual to the azimuthal Killing vector field $\partial/\partial_{\phi}$ and $S$ is the boundary of any spacelike hypersurface; again, for the second equality we have already chosen $S$ to be a spatial section of the event horizon of one of the two black holes. Thus $a_1<z<a_2$ ($a_4<z<a_5$)  for the first (second) black hole. We find
\bequ
J^{Komar}_1=\frac{\pi}{4}
\frac{2a_{51}(a_{42}\tilde{b}-a_{21}\tilde{c})}{a_{41}(2a_{42}a_{51}+b\tilde{c})}\,\Delta
\ , \ \ \ \ \ \
J^{Komar}_2=\frac{\pi}{4}\frac{(\tilde{b}+\tilde{c})}{a_{41}}\,\bar{\Delta}\ . 
\eequ
Thus the angular momentum to mass ratio of any of the individual black
holes has a very simple expression 
\[
j_1\equiv \frac{J_1^{Komar}}{M_1^{Komar}}=
\frac{2}{3a_{41}}\left(\tilde{b}-\frac{a_{21}}{a_{42}}\tilde{c}\right)
\ , \ \ \ \ \ \ \ 
j_2\equiv \frac{J_2^{Komar}}{M_2^{Komar}}=\frac{2}{3a_{41}}\,(\tilde{b}+\tilde{c})\ . 
\]
A simple interpretation for the parameter $c$ follows: it is,
up to a constant, the \textit{difference in angular momentum per unit mass of
the two black holes}
\bequ
c=\frac{3}{2}\frac{a_{42}}{a_{43}}\left(j_2-j_1\right) \label{c}
\ . \eequ
The parameter $b$, on the other hand, is a measure of the \textit{sum}
of the angular momentum per unit mass of the two black holes since
\bequ
b=\frac{3}{2}\left(\frac{a_{42}}{a_{31}}j_1+\frac{a_{21}}{a_{31}}j_2\right) \label{b}
\ . \eequ
Note that, if $c=0$, 
\bequ
j_1=j_2= \frac{2}{3}\frac{a_{31}}{a_{41}}\, b\ .
\eequ
Thus, one should regard $b$ as turning on the angular momentum
per unit mass of both black holes, and one should think of
$c$ as turning on the difference in angular momentum per unit mass of
the two black holes.

One can turn off the intrinsic spin of either black hole by imposing
the conditions
\[ j_1=0 \ \ \Leftrightarrow \ \
\tilde{b}=\frac{a_{21}}{a_{42}}\tilde{c} \ , \ \ \ \ \
j_2=0 \ \ \Leftrightarrow \ \
\tilde{b}=-\tilde{c} \ . \]
One can, however, show that neither these conditions is compatible
with the axis condition (\ref{axiscondition}) and non-trivial $b$ and
$c$. This is most easily done re-expressing the axis condition in
terms of $j_1$ and $j_2$. We get
\bequ
\Delta_{axis}=\frac{3}{2}\frac{a_{41}a_{42}}{a_{43}}\left[2a_{51}(a_{41}j_1-a_{31}j_2)-\frac{9}{4}j_2(j_2-j_1)(a_{42}j_1+a_{21}j_2)\right]
\ . \eequ

The Komar masses and angular momenta, (\ref{komarmass}) and
(\ref{komarj}), obey, together with the temperatures and areas
(\ref{atbh1}) and (\ref{atbh2}), Smarr relations (\ref{smarrgeneral}),
as in the static case and, for instance, the black saturn solution;
but unlike these backgrounds, for our solution the Komar masses and
angular momenta, in general, do not add up to the ADM mass and angular
momentum, since
\bequ
M_{ADM}= M_1^{Komar}+M_2^{Komar}+M^{Komar}_{extra} \ , 
\eequ
\bequ
J_{ADM}= J_1^{Komar}+J_2^{Komar}+J^{Komar}_{extra} \ . \eequ
The reason is that, in general, there is a non-trivial
Komar integral coming from the surface $S_{\phi}$, which is given by
$\rho=0$, $a_3<z<a_4$. This contribution is only present if the axis
condition is not obeyed and it accounts for the extra piece in the last two equations:
\[
M^{Komar}_{extra}=\frac{3}{32\pi G_5}\int_{S_{\phi}} dz
d\phi d\psi
\frac{g_{\rho\rho}g_{\psi\psi}}{\sqrt{-g}}\left[g_{t\phi}g_{t\phi,\rho}-g_{\phi\phi}g_{tt,\rho}\right]=-\frac{3\pi}{8}\frac{a_{43}b\Delta_{axis}}{a_{41}a_{51}(2a_{42}a_{51}+b\tilde{c})}
\ , \]
\[
\barr{l}
\displaystyle{J^{Komar}_{extra}=-\frac{1}{16\pi G_5}\int_{S_{\phi}} dz d\phi d\psi
\frac{g_{\rho\rho}g_{\psi\psi}}{\sqrt{-g}}\left[g_{t\phi}g_{\phi\phi,\rho}-g_{\phi\phi}g_{t\phi,\rho}\right]}\\
\displaystyle{~~~~~~~~~~ =-\frac{a_{43}\Delta_{axis}}{3a_{51}a_{41}^2
  a_{42}}\left(\frac{3\pi}{4}a_{42}+M_1^{Komar}\right) \ .} \earr \]
Note that the extra piece is indeed proportional to $a_{43}$. Imposing
  the axis condition, the Komar masses and angular momenta do add up to the
  ADM mass and angular momentum.

%The reason is that, since in general $\Delta_{axis}\neq 0$, there is a non-trivial
%Komar integral coming from the surface with closed timelike curves 
%$S_{ctc's}$, associated to the rod at $\rho=0$, $a_3<z<a_4$, which
%accounts for the extra piece in the last two equations:
%\[
%M^{Komar}_{ctc's}=\frac{3}{32\pi}\int_{S_{ctc's}} dzd\phi d\psi\,
%\frac{g_{\rho\rho}g_{\psi\psi}}{\sqrt{-g}}\left[g_{t\phi}g_{t\phi,\rho}-g_{\phi\phi}g_{tt,\rho}\right]=-\frac{3\pi}{8}\frac{a_{43}\,b\,\Delta_{axis}}{a_{41}a_{51}(2a_{42}a_{51}+b\tilde{c})}
%\ , \]
%\[J^{Komar}_{ctc's}=-\frac{1}{16\pi}\int_{S_{ctc's}} dz d\phi d\psi\,
%\frac{g_{\rho\rho}g_{\psi\psi}}{\sqrt{-g}}\left[g_{t\phi}g_{\phi\phi,\rho}-g_{\phi\phi}g_{t\phi,\rho}\right]=-\frac{a_{43}\Delta_{axis}}{3a_{51}a_{41}^2
%  a_{42}}\left(\frac{3\pi}{4}a_{42}+M_1^{Komar}\right) \ . \]
%Note that the extra piece is indeed proportional to $a_{43}$. 

\section{Discussion and Conclusions}
In this paper we have used the inverse scattering technique to
generate a new asymptotically flat, vacuum solution of five
dimensional general relativity describing two Myers-Perry black holes,
each with a singular angular momentum parameter, both in the same
plane. We have described the basic properties and physical quantities
of the solution as well as of the background geometry it is built
upon. In general the solution has conical
singularities in both spatial 2-planes. The conical singularity in the
$\rho - \psi$ plane can be removed if 
\[b\tilde{c}=2a_{42}a_{51}\left(\frac{a_{41}a_{52}}{\sqrt{a_{51}a_{31}a_{32}a_{42}}}-1\right)
. \]
On the other hand, the conical singularity in the $\phi$ plane cannot be removed. Indeed, when the axis condition is imposed, which guarantees that $\rho=0$, $a_3<z<a_4$ is an axis, $\delta_{\phi}\neq 0$. The axis condition, which has been interpreted as a torque balance condition is
\[
(\tilde{b}+\tilde{c})(2a_{42}a_{51}-\tilde{b}c)=2a^2_{41}a_{51}c \
. \]
It would be interesting to have a physical interpretation of these
conditions in terms of the different forces and torques that play a role
in this geometry. This might be possible to do using an energetics analysis along
the lines of \cite{Costa:2000kf,gibmal}, a problem we expect to address in the 
future.

One somewhat unexpected feature that we found was a contribution to
the ADM mass and angular momentum of one part of the geometry exterior
to the black hole horizons, if the axis condition is not obeyed. This
suggests that, in the post-post Newtonian analysis of this type of problems, 
along the lines of \cite{Bonnor:2001mh,Bonnor:2001}, one should indeed include
one further parameter describing the rotating rod, as suggested in
\cite{Manko:2003jq}. This might clarify the discrepancy between the result 
obtained in the post-post Newtonian analysis and the one obtained from the exact double Kerr 
solution, for the regularity and axis conditions in the case of two massive 
spinning particles in $D=4$. 

Finally let us remark that in the five dimensional family of
supersymmetric multi-black hole spacetimes known as BMPV 
\cite{Breckenridge:1996is,Gauntlett:1998fz},  no condition is required, 
analogous to the axis condition that has
to be imposed for the IWP spacetimes. This is in curious contrast with the
smoothness properties of horizons in static multi-centre solutions, pointed out in
\cite{Welch:1995dh,Candlish:2007fh}, which get worse in five than in four
dimensions and still worse in higher dimensions.

\section*{Acknowledgements}
We are very grateful to Pau Figueras for correspondence and help in
understanding some aspects of the black saturn solution. We are also
very grateful to  Gary Gibbons, Harvey Reall and Wafic Sabra for discussions
concerning the solution presented herein. C.H. wishes to thank the
hospitality of the D.A.M.T.P. of the University of Cambridge, where
this work was finished. C.R. is funded by FCT through grant SFRH/BD/18502/2004.
This work was also funded by the FCT-CERN grant POCI/FP/63904/2005. Centro de F\'\i sica do Porto is partially funded by FCT throught the POCI programme.

\end{document}